\documentclass[onecolumn, superscriptaddress, preprint, 12pt, letterpaper]{revtex4}
\usepackage{amstext}
\usepackage{amsmath}
\usepackage{amssymb}
\usepackage{graphicx}
\usepackage{psfrag}

\newcommand{\ud}{\mathrm{d}}
\newcommand{\vect}[1]{\boldsymbol{#1}}
\newcommand{\eq}[1]{Eq.~\eqref{#1}}
\newcommand{\fig}[1]{Fig.~\ref{#1}}
\newcommand{\stn}[1]{Sec.~\ref{#1}}
\newcommand{\be}{\begin{equation}}
\newcommand{\ee}{\end{equation}}
\newcommand{\ba}{\begin{align}}
\newcommand{\ea}{\end{align}}
\newcommand{\ti}[1]{\text{#1}}
\newcommand{\mc}[1]{\mathcal{#1}}
\newcommand{\mean}[1]{\langle #1 \rangle}
\newcommand{\ket}[1]{| #1 \rangle}
\newcommand{\bra}[1]{\langle #1 |}
\newcommand{\w}{\omega}

\begin{document}
\title{Electronic coherence dynamics in \emph{trans}-polyacetylene oligomers}
\author{Ignacio Franco}
\altaffiliation[Present address: ]{Theory Department, Fritz Haber Institute of the Max Planck Society, Faradayweg 4-6, 14195 Berlin, Germany.} 
\email{franco@fhi-berlin.mpg.de}
\affiliation{Department of Chemistry, Northwestern University, Evanston, Illinois 60208-3113}

\author{Paul Brumer}
\email{pbrumer@chem.utoronto.ca}
\affiliation{Chemical Physics Theory Group and Department of Chemistry, University of Toronto, Toronto, Ontario, Canada M5S 3H6}

\date{\today}
\begin{abstract}
Electronic coherence dynamics  in \emph{trans}-polyacetylene oligomers are 
considered by explicitly computing the time dependent molecular polarization from the coupled dynamics of electronic and vibrational degrees of freedom in a mean-field mixed quantum-classical approximation. The oligomers are described by the 
Su-Schrieffer-Heeger Hamiltonian and the effect of decoherence is incorporated by propagating an ensemble of quantum-classical trajectories with initial conditions obtained by sampling the Wigner distribution of the nuclear degrees of freedom. 
The electronic coherence of superpositions between the ground and excited and between 
pairs of excited  states is examined for chains of different length, and the dynamics is 
discussed in terms of the nuclear overlap function that appears in the off-diagonal elements of the electronic reduced density matrix. For long oligomers the loss of coherence occurs in tens of femtoseconds. This timescale is determined 
by the  decay of population into other electronic states 
through vibronic interactions, and is relatively insensitive to the type and class of superposition considered. By contrast, for smaller oligomers the decoherence timescale depends strongly on the initially selected superposition, with superpositions that can decay as fast as 50 fs and as slow as 250 fs. The long-lived superpositions are such that little population is transferred to other electronic states and for which the vibronic dynamics is relatively harmonic.

\end{abstract}

\maketitle

\section{Introduction}

Electronic decoherence  (the decay of the off-diagonal elements of the electronic reduced density matrix) in molecules is a basic feature of the electron-vibrational evolution that accompanies  photoexcitation~\cite{franco08}, passage through conical intersections~\cite{martinez2007}, energy transfer~\cite{fleming09} or any other dynamical process that creates electronic superposition states.
 In the decoherence language~\cite{breuer}, the electrons are the system of interest, the nuclei act as the bath, and it is the system-bath interaction what leads to the decoherence.
 Establishing mechanisms for electronic decoherence is central to our understanding of the dynamics underlying fundamental processes such as photosynthesis, vision or electron transport~\cite{martinez2007, fleming09, choi2010}. It is also vital in the development of approximation schemes to the full vibronic evolution of molecules~\cite{kapral06, joe2011}, and it is the starting point for the design of methods to preserve the coherence of electronic superpositions in molecules that can be subsequently exploited in intriguing and potentially useful ways via quantum 
control~\cite{pbnewbook} or quantum information~\cite{nielsen} schemes.

\begin{figure}[htbp]
\begin{center}
\includegraphics[width=0.4\textwidth]{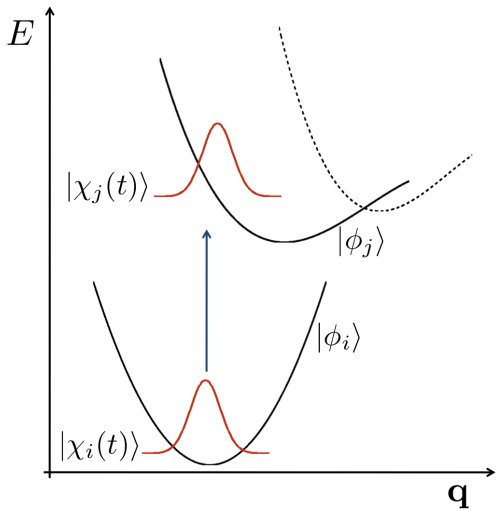}
\caption{Evolution and decay of the overlap of the nuclear wavefunctions in two electronic potential energy surfaces  $S_{ij}(t) = \mean{\chi_j(t)|\chi_i(t)}$  upon instantaneous excitation from state $\ket{\phi_i}$ to state $\ket{\phi_j}$.  Both anharmonicities and population transfer to other electronic states (dotted lines) can lead to a decay of $S_{ij}(t)$ and thus to decoherences between $\ket{\phi_i}$ and $\ket{\phi_j}$.  In the scheme, $E$ is the energy and $\bf{q}$ denotes the general nuclear conformational space. }
\label{fig:decoherence}
\end{center}
\end{figure}

Timescales for electronic decoherence in polyatomic molecules are often exceedingly fast, on the order of tens of femtoseconds~\cite{hwang04, prezdho06, prezdho07, franco08}. This timescale is normally determined by the vibrational degrees of freedom of the nuclear dynamics, with slower torsional, rotational or possible solvent dynamics (if present)  playing a secondary role~\cite{hwang04}. However, challenges in understanding electronic decoherence have arisen from recent spectroscopic observations that have demonstrated that in some photosynthetic systems  electronic coherences can be long-lived~\cite{engel07, lee07, collini09}, with lifetimes exceeding 400-600 fs. These results have lead to discussions of the role of quantum coherences in biological processes and reconsideration of our understanding of decoherence dynamics in single molecules and molecular 
aggregates~\cite{fleming09, engel07, lee07, collini09, Mohseni2008, Lloyd2009,akihito09, briggs2011,
Katz2008,nico2011, Ishizaki2009,kelly2011,leonardo}.  
Many of the associated computations utilize 
phenomenological models or master equations~\cite{pachonpccp} that approximate
the dynamical effects of the bath on the system coordinates
without explicitly following the bath dynamics. In these
approaches, the effect of the bath on the dynamics is
typically determined by adjustable parameters that can be
chosen to reproduce experimental findings, when available.
When possible, however, explicitly following the dynamics of
the nuclei is much preferred~\cite{coker, burghardt2011}. This is because
electronic decoherence in molecules can be understood as
arising  from nuclear dynamics on several electronic
potential   energy   surfaces~\cite{hwang04,  prezdho06,
prezdho07, franco08}. For example, for an entangled vibronic
state of two levels, of the form
\be
\label{eq:vibronic}
\ket{\Psi(t)} = \ket{\phi_i}\ket{\chi_i(t)} +  \ket{\phi_j}\ket{\chi_j(t)} \quad (i\ne j),
\ee
where the $\ket{\phi_n}$ are orthonormal electronic states and  $\ket{\chi_n(t)}$ denotes the nuclear state in the $n^{th}$ electronic surface, the electronic reduced density matrix 
$\rho_e$ is given by:
\be
\begin{split}
\label{eq:vibronicrho}
\rho_e(t) & =  \textrm{Tr}_\ti{N}\{ \ket{\Psi(t)}\bra{\Psi(t)}\}  \\
& =  \ket{\phi_i}\bra{\phi_i} \mean{\chi_i (t) | \chi_i(t)} +  \ket{\phi_j}\bra{\phi_j}\mean{\chi_j (t) | \chi_j(t)}  + \left[\ket{\phi_i}\bra{\phi_j} \mean{\chi_j (t) | \chi_i(t)} + \textrm{h.c.}\right]~.
\end{split}
\ee
Here the trace is over the nuclear states and  h.c. denotes the  hermitian conjugate.  Hence, the decay of the off-diagonal matrix elements in $\rho_e(t)$, i.e., electronic decoherence, is governed by the degree of overlap of the nuclear wavepackets  $S_{ij}(t) =\mean{\chi_j(t) | \chi_i(t)}$ associated with the electronic states in the superposition. Thus, by understanding the events that lead to a decay of the overlaps $S_{ij}(t)$ one  obtains direct insights into the mechanism of electronic decoherence between states $i$ and $j$. A schematic representation of such evolution and decay for a particular pair of states is shown in \fig{fig:decoherence}.

In this paper we present a study of the electronic coherence dynamics in \emph{trans}-polyacetylene (PA) oligomers in which the dynamics of both electronic and vibrational degrees of freedom are explicitly taken into account.  
We do so in an approximate scheme where the nuclei are considered classically and the electrons quantum mechanically.  The oligomers are described using the well-known Su-Schrieffer-Heeger (SSH) Hamiltonian~\cite{SSH}. The SSH model treats the molecule as a tight-binding chain in which the electrons are coupled to distortions in the polymer backbone by electron-vibrational interactions. 
In spite of its simplicity, the SSH Hamiltonian is remarkably successful in capturing the electronic structure of PA, its photoinduced vibronic dynamics and the rich photophysics of polarons, breathers and kinks~\cite{kobayashi2009, kobayashi2002, franco08,sergei2003}. This model is often used to study the dynamical features caused by strong electron-ion couplings~\cite{franco08,stella2011, ness1999}.  

The coupled dynamics of nuclear and electronic degrees of freedom of the molecule is followed in a mean-field (Ehrenfest) mixed quantum-classical approximation~\cite{tully, streitwolf1, johansson1} and decoherence effects are incorporated by propagating an ensemble of quantum-classical trajectories with initial conditions selected from the nuclear Wigner distribution function~\cite{franco08,francowire, francowireprl, wigner} of the chain.  In this way the dynamics reflects the initial nuclear quantum distribution and is subject to the level broadening and internal relaxation mechanism induced by the vibronic couplings.
Using this model we study the possible effect of system size,  nuclear initial conditions and type of electronic superposition states on the  dynamics of electronic coherence.

\section{Model and Methods}
\label{stn:methods}

\subsection{The SSH Hamiltonian}

The SSH Hamiltonian~\cite{SSH} models PA oligomers  as one-dimensional tight-binding chains, each site representing a CH unit. The Hamiltonian for an $N$-membered  oligomer
is given by:
\be
H_\ti{SSH} = H_\ti{elec} + H_\ti{ph},
\ee
where
\be
\begin{split}
H_\ti{elec} &= \sum_{n=1}^{N-1} \sum_{s=\pm1}[-t_0 +\alpha(u_{n+1}-u_n)](c_{n+1,s}^\dagger c_{n,s} + c_{n,s}^\dagger c_{n+1,s}) \quad \ti{and,} \\
H_\ti{ph} &= \sum_{n=1}^{N} \frac{p_n^2}{2M} + \frac{K}{2}\sum_{n=1}^{N-1}(u_{n+1}-u_n)^2,
\end{split}
\ee
are, respectively, the electronic and nuclear parts of the Hamiltonian. Here, $u_n$ denotes the displacement of the $n$th CH site from the perfectly periodic position $x=na$ with $a$ as the lattice constant,  $M$ is the mass of the CH group, $p_n$ is the momentum conjugate to $u_n$ and $K$ is an effective spring constant. The operator  $c_{n,s}^\dagger$ (or $c_{n,s}$) creates (or annihilates) a fermion 
on site $n$ with spin $s$ and satisfies the usual fermionic anticommutation relations. The electronic component of the Hamiltonian consists of a term describing the hopping of $\pi$ electrons along the chain with hopping integral $t_0$ and an electron-ion interaction term with coupling constant $\alpha$.
The quantity $\alpha$ couples the  electronic states to the molecular geometry and constitutes a first-order correction to the lowest-order hopping integral $t_0$.     Throughout this work, we use the standard set of SSH parameters for PA: $t_0=2.5$ eV, $\alpha=4.1$ eV/\AA, $K=21$ eV/\AA$^2$, $M=1349.14$ eV fs$^2$/\AA$^2$, and $a=1.22$ \AA. 

\subsection{Ehrenfest electron-vibrational dynamics}

The electron-vibrational dynamics of the chain is followed in the mean-field Ehrenfest mixed quantum-classical approximation~\cite{streitwolf1, johansson1, franco08, francowire, francowireprl}. In this approximation, the nuclei move classically on a mean-field potential energy surface with forces given by
\be
\dot{p}_n = - \bra{\varphi(t)} \frac{\partial H_\ti{SSH}}{\partial u_n} \ket{\varphi(t)}.
\ee
In turn,   the antisymmetrized $\mc{N}$ electron wavefunction $\ket{\varphi(t)}$ satisfies the time-dependent Schr\"odinger equation
\be
\label{eq:se}
i\hbar \frac{\partial}{\partial t}\ket{\varphi(t)} = H_\ti{SSH}[\vect{u}(t)]
\ket{\varphi(t)},
\ee
where $\vect{u}\equiv(u_1, u_2, \cdots, u_N)$.  Since $H_\ti{SSH}$ is a single-particle operator, the electronic properties of the system are completely characterized by the single-particle electronic density matrix
\be
\rho_{n,m}(t) = \sum_s \bra{\varphi(t)}c_{n,s}^\dagger c_{m,s}\ket{\varphi(t)}.
\ee
From \eq{eq:se} it follows that the dynamics of $\rho_{n,m}$ satisfies
\be
\label{eq:rdmeom}
i\hbar \frac{\ud}{\ud t} \rho_{n,m}(t) = \sum_s \bra{\varphi(t)}[c_{n,s}^\dagger c_{m,s}, H_\ti{elec} ] \ket{\varphi(t)} = \sum_{m'} (h_{m,m'} \rho_{n,m'}(t) - h_{m',n}\rho_{m',m}(t)),
\ee
where $h_{n,m} = \bra{n,s}H_\ti{elec}\ket{m,s}$ are the single-particle matrix elements of $H_\ti{elec}$ and $\ket{n,s}= c_{n,s}^\dagger \ket{0}$ where $\ket{0}$ is the vacuum state.

Equation \eqref{eq:rdmeom}  is integrated by decomposing $\rho_{n,m}(t)$ into orbitals. For this, let $\ket{\epsilon,s}$ be the eigenorbitals of spin $s$ and energy $\epsilon$ of the system at preparation time ($H_\ti{elec}(t=0) \ket{\epsilon,s} = \epsilon \ket{\epsilon,s}$).
Using this basis,  the initial electronic reduced density matrix  can be expressed as
\be
\label{eq:rdminitial}
\rho_{n,m}(0)
= \sum_{\epsilon, \epsilon'=1}^{N}\sum_{s} \bra{\epsilon,s}n,s\rangle \langle m,s\ket{\epsilon',s} \bra{\varphi(0)} c_{\epsilon,s}^\dagger c_{\epsilon',s} \ket{\varphi(0)}
\ee
 where $\bra{\varphi(0)} c_{\epsilon,s}^\dagger c_{\epsilon',s} \ket{\varphi(0)}$
  characterizes the initial electronic distribution among the single particle states, and $\ket{\epsilon,s}= c_{\epsilon,s}^\dagger\ket{0}$.  In  writing  \eq{eq:rdminitial}  we have employed the basis transformation function
$c_{n,s}^\dagger  = \sum_{\epsilon=1}^{N} \bra{\epsilon,s} n,s\rangle c_{\epsilon,s}^\dagger$.
We adopt the ansatz  that upon time evolution   $\rho_{n,m}(t)$ maintains the form in \eq{eq:rdminitial}. That is,
\be
\label{eq:rdmtime}
\rho_{n,m}(t) =
\sum_{\epsilon, \epsilon'=1}^{N}\sum_{s} \bra{\epsilon(t),s}n,s\rangle \langle m,s\ket{\epsilon'(t),s} \bra{\varphi(0)} c_{\epsilon,s}^\dagger c_{\epsilon',s} \ket{\varphi(0)}.
\ee
The utility of this ansatz is that if the time-dependent
orbitals $\ket{\epsilon(t),s}$ satisfy the single-particle Schrodinger equation
\be
\label{PAeq:electmp}
 i\hbar \frac{\ud}{\ud{t}}\ket{\epsilon(t),s}
=   H_\ti{elec}(t)\ket{\epsilon(t),s},
\ee
with initial conditions $ \ket{\epsilon(t=0),s} =
\ket{\epsilon,s}$, the reduced density matrix  automatically
satisfies the correct equation of motion [\eq{eq:rdmeom}].

Within this framework, the equations for the nuclear trajectories are:
\begin{equation}
\label{PAeq:nuclei}
\begin{split}
\dot{u}_n(t)  =& \frac{p_n(t)}{M}; \\
\dot{p}_n(t)  =&  - K\left(2u_n(t) - u_{n+1}(t) - u_{n-1}(t)\right)
+ 2\alpha\textrm{Re}\left\{ \rho_{n, n+1}(t)
-   \rho_{n, n-1}(t) \right\}.
\end{split}
\end{equation}
 The chain is taken to be clamped so that $u_1(t)=u_N(t)=0$ and
$p_1(t)=p_N(t)=0$ for all time, and \eq{PAeq:nuclei}  is valid for
$n=2, \cdots, N-1$. In turn, the orbitals that form
$\rho_{nm}(t)$ satisfy \eq{PAeq:electmp}, so that
\be
\label{PAeq:elec}
\begin{split}
 i\hbar \frac{\ud}{\ud t} \langle n \ket{\epsilon(t)}
 = & \left[-t_0 + \alpha (u_{n+1}(t)-u_n(t))\right]\langle n+1\ket{\epsilon(t)} \\
+ & \left[-t_0 + \alpha (u_{n}(t)-u_{n-1}(t))\right]\langle{ n-1}\ket{\epsilon(t)}
\end{split}
\ee
for $n, \epsilon=1, \cdots, N$. Since the electrons are confined
within the chain,  $\bra{n}\epsilon(t)\rangle=0$ for $n \notin
\{1, \cdots, N\}$. Equations~\eqref{PAeq:nuclei}
and~\eqref{PAeq:elec} constitute a closed set of $N(N+2)$
coupled first-order differential equations that are integrated
using  an eighth-order  Runge-Kutta method.

\subsection{Nuclear initial conditions}
\label{stn:ninit}

For the purpose of determining the nuclear initial conditions, the electronic state $\ket{\varphi(0)}=\ket{E_0}$ ($H_\ti{elec} \ket{E_0} = E_0\ket{E_0}$) is assumed to be well described by a single Slater determinant for which [recall \eq{eq:rdminitial}]
\be
\bra{\varphi(0)} c_{\epsilon,s}^\dagger c_{\epsilon',s} \ket{\varphi(0)} = \delta_{\epsilon, \epsilon'} f(\epsilon, s),
\ee
where $f(\epsilon,s)$ is the initial electronic distribution (ground or excited) that takes values 0 or 1 depending on the initial occupation  of each level with energy $\epsilon$ and spin $s$. The starting optimal 
(minimum energy) geometry is obtained by minimizing the total
energy of the chain  by an iterative self-consistent procedure.  Specifically, the  energy gradient  of the oligomer is given by
\be
\label{PAeq:grad2}
\begin{split}
\frac{\partial E (\vect{u})}{\partial u_m} =
\bra{\varphi(0)} \frac{\partial H}{\partial u_m} \ket{\varphi(0)}
=
2\alpha \text{Re}\{\rho_{m, m-1} -\rho_{m, m+1}  \}
  + K(2u_m - u_{m-1} - u_{m+1}).
\end{split}
\ee
At the optimal geometry, for which the gradient equals zero, the $m=2, \ldots, N-1$ displacement satisfies
\be\label{PAeq:iter}
u_{m}   =  \frac{1}{2}\left(u_{m+1} + u_{m-1}\right)
- \frac{\alpha}{K} \text{Re}\{\rho_{m, m-1} -\rho_{m, m+1}  \}.
\ee
Equation \eqref{PAeq:iter} is solved iteratively  with the additional constraint that  the boundaries of the chain are  clamped ($u_1 = u_N = 0$).

Subsequently, a harmonic approximation to the nuclear ground-state wavefunction is obtained by performing a normal mode analysis around the equilibrium minimum energy geometry $\vect{u}^{0}= (u_1^{0}, \cdots, u_N^{0})$ in the (ground or excited) initial electronic state $\ket{E_0}$.  For this,  the Hamiltonian  is expressed as  a
sum of the  static equilibrium configuration
$H_0$ and a dynamical part due to deviations from equilibrium
\begin{equation}
H =   H_{0} + H_{\pi-\rm{ph}}' + H_{\rm{ph}}',
\end{equation}
where
\begin{align*}
H_{\pi-\rm{ph}}' &= \alpha\sum_{n=1, s}^{N-1} (\eta_{n+1} - \eta_n)
(c_{n+1,s}^{\dagger}c_{n,s} + c_{n,s}^{\dagger}c_{n+1,s})\\
\displaybreak[0]
H_{\rm{ph}}'  =   \sum_{n=1}^{N} \frac{p_n^2}{2M} & + \frac{K}{2}
\sum_{n=1}^{N-1}\Big[ 2(u_{n+1}^0 - u_n^0)(\eta_{n+1} - \eta_n)
+ (\eta_{n+1}-\eta_n)^2 \Big],
\end{align*}
with $\eta_n$ being the displacement of the $n^{th}$ monomer from its equilibrium
position $\eta_n= u_n-u_n^0$.  In order to get the potential energy of the chain around the equilibrium geometry, the quantity
$H_{\pi-\rm{ph}}'$ is considered as a perturbation to $H_0$~
\cite{phonon1, phonon2} and we have to second order that,
\be
\label{PAeq:pot}
\begin{split}
E(\vect{\eta})  & =  E_0  + \langle E_0 | H_{\pi-\rm{ph}}'
| E_0 \rangle
+ \sum_{i\ne 0} \frac{|\langle \varphi_i | H_{\pi-\rm{ph}}' | E_0\rangle
| ^2}{E_0 - E_i} \\
  &+ \frac{K}{2}\sum_{n=1}^{N-1}[ 2(u_{n+1}^0 - u_n^0)(\eta_{n+1}-\eta_n)
+ (\eta_{n+1} - \eta_n)^2 ]
\end{split}
\ee
where we have traced over the electronic coordinates and  assumed that the
system is initially prepared in the electronic state $\ket{E_0}$
with energy $E_0$. Here $\{|\varphi_i\rangle, E_i\}$ are the eigenstates and
eigenvalues of the $\mathcal{N}$-particle electronic Hamiltonian in the optimal
geometry $H_0$.

A harmonic version of Eq.~\eqref{PAeq:pot} is obtained by making a
Taylor expansion of the potential around the equilibrium position and keeping
terms up to second order in the nuclear displacements.  We note that
second-order perturbation in $H_{\pi-\rm{ph}}$ is consistent with the harmonic
approximation. The effective harmonic phonon potential energy thus obtained is:
\begin{subequations}
\label{PAeq:harmonizedchain}
\be
E^{\text{harm}}(\vect{\eta}) =  E_0  +  \frac{1}{2}\sum_{n,m=2}^{N-1}
\eta_n f_{n,m}\eta_m.
\ee
Here $f_{n,m}$ is the Hessian of the potential energy given by:
\be
\begin{split}
f_{n,m} & = \frac{\partial^2 E}{\partial \eta_n\, \partial \eta_m}
\bigg|_{\vect{\eta}=\vect{0}}
= V_{nm} + K(2\delta_{n,m} -\delta_{n,m+1} -\delta_{n, m-1}),
\end{split}
\ee
where
\begin{equation}
\begin{split}
V_{nm}   & = 2\alpha^2 \sum_{\epsilon, \epsilon', s} \frac{ f(\epsilon',s)
(1-f(\epsilon,s))}{\epsilon' - \epsilon} V^m(\epsilon, \epsilon') V^n(\epsilon, \epsilon') ,\\
V^n(\epsilon, \epsilon') = \langle\epsilon|n\rangle  &
\Big(\langle n-1|\epsilon'
\rangle - \langle n+1|\epsilon'\rangle\Big)
+ \langle n | \epsilon'\rangle\Big( \langle \epsilon| n-1\rangle - \langle
\epsilon | n+1 \rangle\Big).
\end{split}
\end{equation}
\end{subequations}
In deriving \eq{PAeq:harmonizedchain} we have   imposed clamped ends on the
polymer chain ($\eta_1=\eta_N=0$). The orbitals $\ket{\epsilon}$ and their
associated energies $\epsilon$ are obtained by diagonalizing the electronic
Hamiltonian at the equilibrium geometry.  The normal mode coordinates and
frequencies are then computed by the standard analysis~\cite{wilson}. The
eigenvectors of  $f_{nm}$ provide the normal mode coordinates
$Q_j(\vect{\eta})$ and the associated eigenvalues $\lambda_j$ the normal mode
frequencies $\w_j = \sqrt{\lambda_j/M}$.

A phase-space like description of the resulting nuclear quantum state is obtained by constructing  the associated nuclear Wigner phase-space distribution function $\rho_\ti{W}(\vect{u}, \vect{p})$. In the normal-mode coordinates,   $\rho_\ti{W}(\vect{u}, \vect{p})$  is  just the product of the Wigner distributions associated with each vibrational
mode
\be
\label{PAeq:wignerlattice}
\rho_\ti{W}(\vect{u}, \vect{p}) = \prod_{j=1}^{N-2} \rho_j(Q_j(\vect{u}),
P_j(\vect{p})),
\ee
where  $Q_j(\vect{u})$ is the normal mode coordinate of the $j$-th mode and
$P_j(\vect{p})$ its  conjugate momentum. We take the chain to be initially
prepared in its ground vibrational state so that~\cite{wigner}
\be
\rho_j(Q_j, P_j) =\frac{1}{\pi\hbar} \exp({- M \w_j Q_j^2/\hbar})
\exp({-P_j^2/\hbar\w_j M})
\ee
for $j=1,\cdots, N-2$.  The $2N-4$ dimensional phase-space distribution in
\eq{PAeq:wignerlattice} completely characterizes the initial quantum state of
the nuclei.

The ensemble of lattice initial conditions,  $\{\vect{u}^i(0), \vect{p}^i(0)\},$  for the quantum-classical dynamics is obtained from a   Monte Carlo  sampling of the nuclear Wigner phase space distribution of \eq{PAeq:wignerlattice}.
The average classical energy of the resulting ensemble
coincides numerically with the zero-point energy of the lattice. The associated initial values for the orbitals $\{ \ket{\epsilon^i}\}$ are obtained by diagonalizing $H_\ti{elec}$ in the initial lattice geometries $\{\vect{u}^i\}$. Each initial condition $i$, together with the equations of motion, defines a quantum-classical trajectory $(\vect{u}^i(0), \vect{p}^i(0), \ket{\varphi^i(0)}) \rightarrow  (\vect{u}^i(t), \vect{p}^i(t), \ket{\varphi^i(t)})$ and the set is employed to obtain ensemble averages. Results shown here are averages over 10000 trajectories.

\section{Results and discussion}

\begin{figure}[htbp]
\begin{center}
\includegraphics[width=0.6\textwidth]{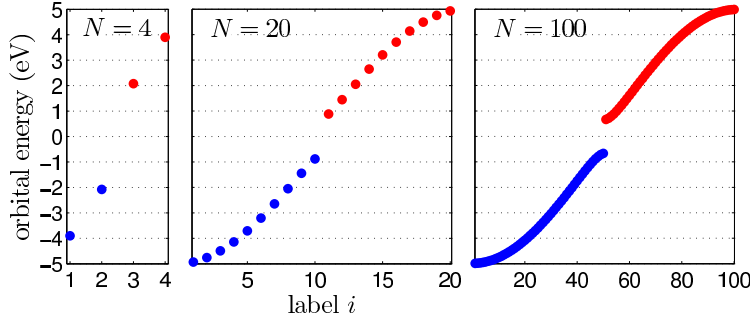}
\caption{Orbital energies for optimized PA chains of varying length. The valence (conduction) orbital energies are plotted in blue (red).  }
\label{fig:initial}
\end{center}
\end{figure}

Throughout, we study neutral oligomers with an even number $N$ of CH units. In the ground state, the geometry of the chain consists of a centrosymmetric structure with perfect alternation of double and single bonds.
 The single-particle spectrum of chains of different length is shown in \fig{fig:initial}.
 It has a total width of $4t_0=10$ eV and consists of $N/2$ fully occupied valence states and $N/2$ initially empty conduction states. Note how the single-particle spectrum gets more dense as the number of CH units  is increased.

As a measure of electronic coherence and decoherence we follow the dynamics of the  molecular polarization, defined by $\langle \hat{\mu}(t) \rangle = \bra{\Psi(t)} \hat{\mu} \ket{\Psi(t)}$,
where $\ket{\Psi}$ denotes the vibronic wavefunction and where the dipole operator $\hat{\mu}= \hat{\mu}_\text{e} + \hat{\mu}_\text{N}$ has both an electronic $\hat{\mu}_\text{e}$ and  a nuclear $\hat{\mu}_\text{N}$ component.  Doing so provides a measure of electronic coherence that is directly related to measurables.

It is advantageous to connect this discussion of decoherence based on the polarization to standard measurements of decoherence
\cite{schlosshauer,pbnewbook}. The density matrix associated with a general entangled vibronic Born-Oppenheimer state of the form $|\Psi(t) \rangle = \sum_n e^{-iE_n t/\hbar} | \varphi_n \rangle | \chi_n(t) \rangle$ is given by
\be
\label{eq:rho}
|\Psi(t)\rangle\langle\Psi(t)| = \sum_{nm} e^{-i\w_{nm}t}|\varphi_n\rangle |\chi_n (t)\rangle \langle\varphi_m|\langle\chi_m (t)|,
\ee
where $\ket{\varphi_n}$  are the electronic eigenstates [$H_\ti{elec}\ket{\varphi_n} = E_n \ket{\varphi_n}$], $\ket{\chi_n(t)}$ the nuclear wavepacket associated with each electronic level and $\w_{nm}= (E_n- E_m)/\hbar$. 
 If our interest is in the electronic degrees of freedom only, then the vibrations are regarded as the environment.  Since we have no interest in the behavior of the environment we trace over these modes to give the density matrix of the electronic subsystem:
\be
\label{eq:rrho}
\rho_e (t) = \sum_{nm} e^{-i\omega_{nm}t} \langle \chi_m (t) | \chi_n (t) \rangle  | \varphi_n\rangle \langle \varphi_m | . 
\ee
Note that the off-diagonal elements of $\rho_e(t)$ are determined by the nuclear overlaps $S_{nm}(t)=\langle \chi_m (t) | \chi_n (t) \rangle $ and the loss of such coherences in $\rho_e(t)$ is a result of the evolution of the $S_{nm}(t)$ due to the vibronic dynamics.  Standard measures of decoherence capture precisely this. 
For example, the purity of such entangled vibronic state is given by
\be
\textrm{Tr}(\rho_e^2(t)) = \sum_{nm} |\langle \chi_m(t)|\chi_n(t)\rangle |^2
\ee
and decays with the overlaps of the nuclear wavepackets in the different electronic surfaces. 

The polarization is also a useful measure of decoherence because its magnitude also depends on the $S_{nm}(t)$.  To see this consider the expression for the polarization for the entangled state in \eq{eq:rho}:
\be
\langle \hat{\mu}(t) \rangle = \sum_n \bra{\chi_n(t)} \mu_N \ket{\chi_n(t)}  + \sum_{n,m}   e^{-i \w_{nm}t}\mu_\text{e}^{mn} \mean{\chi_m (t)|\chi_n (t)},
\ee
where $\mu_{e}^{mn} = \bra{\varphi_m} \hat{\mu}_\text{e} \ket{\varphi_n}$. 
Suppose that the PA chain is  prepared in a spatially symmetric state where the initial nuclear state is invariant under reflection, i.e. $\rho_\ti{W}(-\vect{u}, -\vect{p}) = \rho_\ti{W}(\vect{u}, \vect{p})$. Since there is no symmetry breaking term in the Hamiltonian, this initial symmetry is maintained throughout the dynamics~\cite{francosym} and $\bra{\chi_n} \mu_N \ket{\chi_n}= \mu_\text{e}^{nn} = 0$ for all $n$. Under such conditions, the polarization
\be
\label{eq:polarization}
\langle \hat{\mu} (t) \rangle =  \sum_{n,m\ne n}  e^{-i \w_{nm}t} \mu_\text{e}^{mn} \mean{\chi_m (t)|\chi_n(t)}
\ee
is a direct measure of the off-diagonal matrix elements of the electronic reduced density matrix [cf. \eq{eq:rrho}].  Its evolution and decay directly offers information about the decoherence dynamics. 

Note that in writing \eq{eq:polarization} we have adopted the Franck-Condon approximation where the electronic transition dipole surfaces $\mu_\text{e}^{mn}(\vect{u})$ are assumed to depend 
weakly on the nuclear displacements. However even when this approximation is 
not valid, a decay in $\langle \hat{\mu} (t) \rangle$ will still signal a decay in the nuclear overlaps, albeit modulated by the dependence of the electronic transition dipoles on the 
nuclear coordinates.

Thus, both $\textrm{Tr}(\rho_e^2(t)) $ and $\langle\hat{\mu}(t)\rangle$ are useful measures of decoherence and both decay with the  overlaps of the nuclear wavepackets in different electronic states. The advantage of the polarization over the purity is that it is a physically accessible observable. Its limitation, however, is that  $\langle \hat{\mu}(t)\rangle$ only signals coherences for which $\mu_e^{mn} \ne 0$. So, for instance, coherences between eigenstates of the same parity are 
absent in the polarization even when they would  contribute to the purity.

 In the quantum-classical picture of the dynamics, the polarization is computed as an average of the polarizations recorded for each of the $\mc{M}$ individual trajectories in the ensemble:
\be
\label{eq:polarizationqcc}
\langle \hat{\mu}(t)\rangle = \frac{|e|}{\mc{M}} \sum_{i=1}^{\mc{M}} \sum_{n=1}^{N} x_n^i(t) (1- \rho_{n,n}^i)
\ee
where $x_n^i(t) = (na + u_n^i(t))$ is the position of site $n$ at time $t$ in the $i$th trajectory and $e$ is the electron charge. The first term in \eq{eq:polarizationqcc} comes from the dipole due to the nuclei, while the second one quantifies the electronic contributions.

There are two possible effects that can lead to a decay of the electronic coherences (recall \fig{fig:decoherence}): anharmonicities in the potential and population transfer into other electronic states. More precisely, if there is no population transfer into other states, and the electronic potential energy surfaces are bounded, then anharmonicities in the electronic potential energy surfaces can lead to a spread of the nuclear wavepackets  during evolution and thus to a decay of the nuclear overlap integral $S_{ij} = \mean{\chi_j(t) | \chi_i(t)}$ 
(wavepacket evolution in purely harmonic potentials  lead to periodic recurrences in $S_{ij}$ and thus cannot lead to decoherence). Alternatively, population transfer into other electronic states can lead to decoherence by transferring population to states for which only poor overlaps of the nuclear wavepackets are possible with the states already involved in the superposition.  This poor overlap arises because different electronic potential energy surfaces typically have substantially different gradients and position of their minima in conformational space, leading to diverging evolution of the nuclear wavepackets in the excited state manifold.

It should be noted that, traditionally, studies of the decoherence of a 
superposition state would not typically include loss of population 
from the state, which would be regarded as a relaxation,
rather than decoherence, process. However, this distinction
is meaningful when energy transfer and decoherence time
scales are substantially different, the latter occurring on
much shorter time scales than the former. Here, however, as
shown below, population transfer between states 
occur  rather  quickly,  making  this  subdivision  less
meaningful, and making such contributions quite significant in the
time evolution of $\langle \hat{\mu} (t)\rangle$ and the purity. Here then, we
use the term ``decoherence" to relate to any process that causes loss 
of the coherence of $\langle \hat{\mu} (t)\rangle$.

Below we discuss several examples of decoherence dynamics in PA chains.  The interpretation of the results will be done in a wavepacket language and with the wavepacket picture of \fig{fig:decoherence} in mind, 
even though the computations  are performed in a mixed quantum-classical setting. Such wavepacket evolution  is captured by the quantum-classical dynamics through the time dependence of the orbital energies and populations in the ensemble of trajectories.

\subsection{Decoherence between the ground and first excited state  for chains of different lengths}

\begin{figure}[htbp]
\centering
\includegraphics[width=0.8\textwidth]{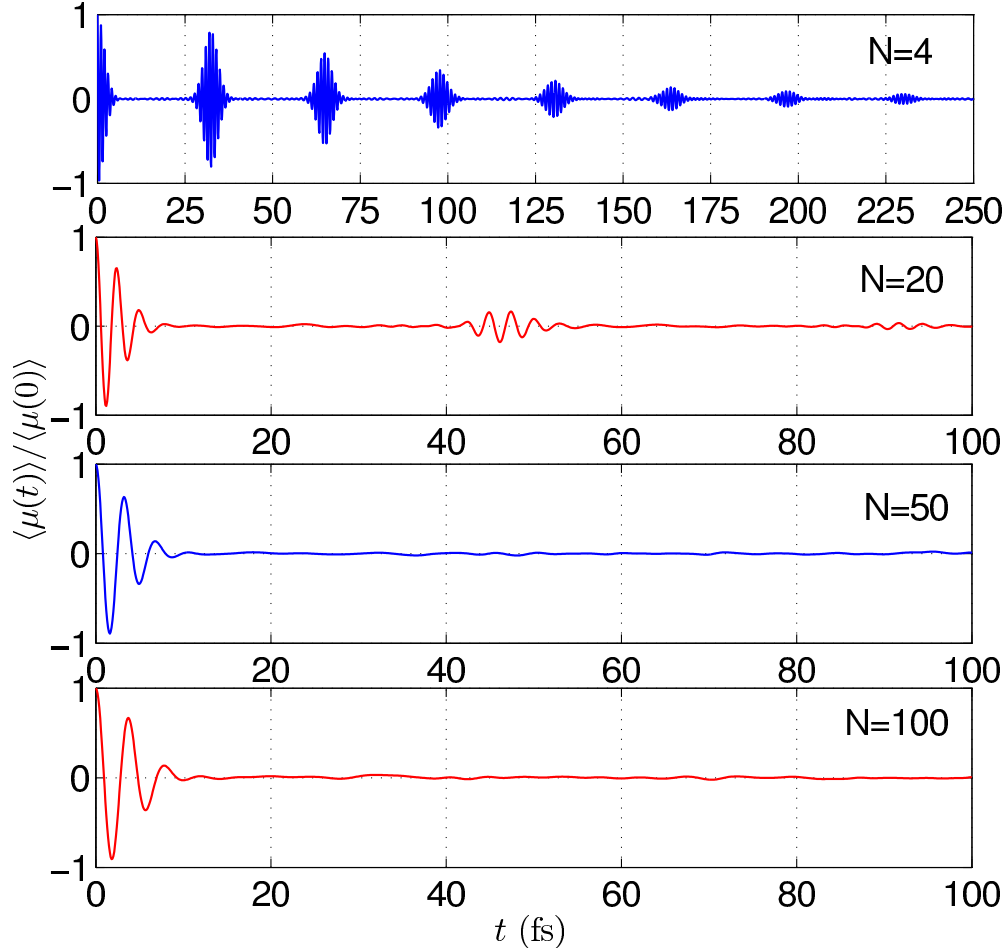}
\caption{Electronic decoherence dynamics in chains of different lengths $N$. The figure shows the evolution and decay of the chain polarization when the system is initially prepared in a superposition between the ground and first excited state 
of the form in \eq{eq:super0}.  }
     \label{fig:GEvsN}
\end{figure}

\begin{figure}[htbp]
\centering
\includegraphics[width=0.5\textwidth]{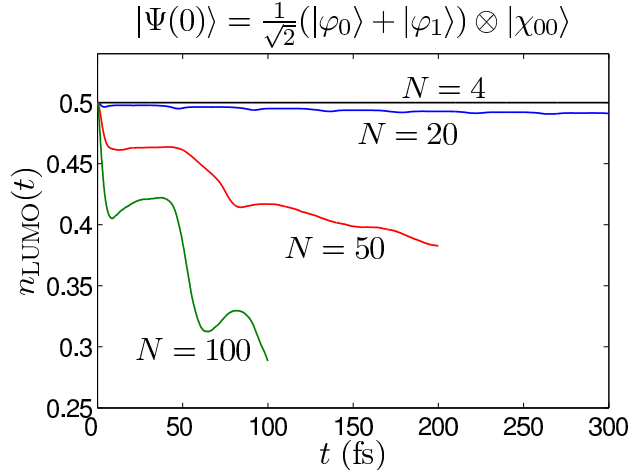}
\caption{Population of the  LUMO orbital during the decoherence dynamics of a N-site SSH chain with the initial conditions of \eq{eq:super0}. }
     \label{fig:poplength}
\end{figure}

Consider first the decoherence dynamics of PA chains initially in a separable superposition state of the form:
\be
\ket{\Psi(0)} = \frac{1}{\sqrt{2}}(\ket{\varphi_0} + \ket{\varphi_1})\otimes\ket{\chi_{00}},
\label{eq:super0}
\ee
where $\ket{\varphi_0}$ is the ground electronic state, $\ket{\varphi_1}$ is the first excited state (obtained by promoting an electron from the HOMO to the LUMO) and  $\ket{\chi_{00}}$ is the ground state nuclear wavefunction in  the ground electronic surface. Physically, such a superposition can be created by instantaneous 
(delta pulse) excitation of the relaxed ground state chain.  Figure~\ref{fig:GEvsN} shows $\mean{\hat{\mu}(t)}$  for chains with varying number of CH units ($N$).   The high frequency oscillations in  $\mean{\hat{\mu}(t)}$ are due to the difference in energy  between the two states involved in the superposition (in this case the energy gap). The remaining time dependence arises from the wavepacket evolution in the excited state   potential energy surface. For the four site chain, the polarization displays a fast initial decay with recurrences every $\sim 30$ fs. These recurrences arise from the time dependence of the overlap of the nuclear wavefunctions in the ground and excited electronic states [see \eq{eq:polarization}], and signal the oscillatory motion of the nuclear wavepacket in the excited state potential. Between consecutive recurrences the amplitude of the polarization diminishes and eventually dies out,  yielding a decoherence timescale of $\sim$ 250 fs for $N=4$.   For $N=20$  we observe only two of these recurrences, occurring every $\sim 46$ fs to yield a decoherence time of $\sim 100$ fs. For longer oligomers  ($N=50$ and $N=100$) no recurrences are observed and the decoherence occurs in less than 10 fs.

Additional insights into the decoherence dynamics can be extracted by considering the evolution of the population of the LUMO of the chain (\fig{fig:poplength}). If the main decoherence mechanism is the anharmonic evolution of the nuclear wavepacket in the first excited state potential energy surface, then one should expect little population exchange with other levels of the chain. As shown in \fig{fig:poplength} for $N=4$ and $N=20$ an almost negligible amount of population is transferred to other electronic states, suggesting that  anharmonicities are the main source of decoherence.  By contrast, for long chains ($N=50$ and $N=100$) the electronic spectrum is so dense that a substantial amount of population is transferred from the initially populated LUMO to other electronic states. This suggests that both anharmonicities and population decay to other electronic states contribute to the decoherence, leading to an evolution with no apparent recurrences.

\subsection{Decoherence of superpositions between excited states}

\begin{figure}[htbp]
\begin{center}
\includegraphics[width=0.6\textwidth]{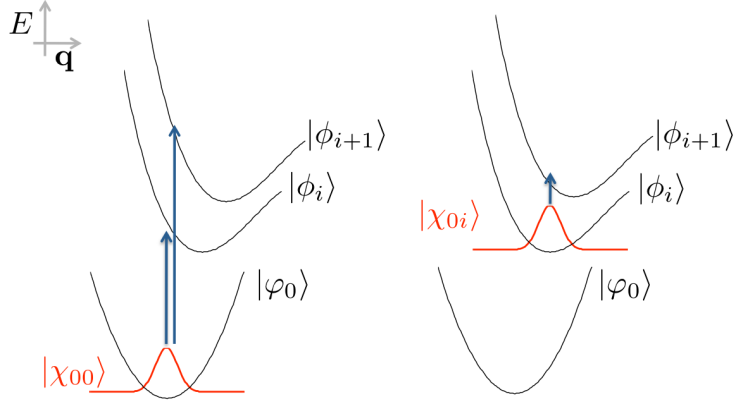}
\caption{Schematic of the process used to create the superpositions in \eq{eq:super1} (left panel) and \eq{eq:super2} (right panel).  }
\label{fig:decoherence2}
\end{center}
\end{figure}

We now investigate how the decoherence dynamics changes when the initial superposition is  between two excited states rather than between an excited and ground electronic state. For this we consider the two classes of 
model initial superpositions schematically represented in \fig{fig:decoherence2}. In the first class, the initial state is of the form
\be
\label{eq:super1}
\ket{\Psi(0)} = \frac{1}{\sqrt{2}}(\ket{\phi_i} + \ket{\phi_{i+1}})\otimes\ket{\chi_{00}},
\ee
where $\ket{\phi_i}  = c_{i,s}^\dagger c_{N/2, s}\ket{\varphi_0}$  ($i\in \{N/2+1, \cdots, N\}$) is an electronically excited state obtained by promoting an electron from the HOMO  to the $i$th orbital level of the ground state $\ket{\varphi_0}$. In this superposition the initial nuclear state is taken to be  the ground vibrational state in the ground electronic surface, $\ket{\chi_{00}}$. Physically, such a superposition will arise 
via instantaneous excitation of the ground  vibronic state into states 
$\ket{\phi_i}$ and $\ket{\phi_{i+1}}$, as depicted in the left panel of \fig{fig:decoherence2}.   By contrast, in the second class of superpositions the nuclei are taken to be initially prepared in the  ground state distribution of the  \emph{excited} electronic state $\ket{\phi_i}$, so that
 \be
 \label{eq:super2}
\ket{\Psi(0)} = \frac{1}{\sqrt{2}}(\ket{\phi_i} + \ket{\phi_{i+1}})\otimes\ket{\chi_{0i}}.
\ee
The wavefunction $\ket{\chi_{0i}}$ is obtained by finding the optimal geometry of the electronically excited state and then performing a normal mode analysis around this geometry, as discussed in \stn{stn:ninit}. Physically such a superposition will arise from instantaneous excitation of a chain vibrationally relaxed in state $\ket{\phi_i}$ to state  $\ket{\phi_{i+1}}$, as represented in the right panel of \fig{fig:decoherence2}.

\subsubsection{The case of a 20-site chain}

\begin{figure}[tp]
\centering
\includegraphics[width=1.0\textwidth]{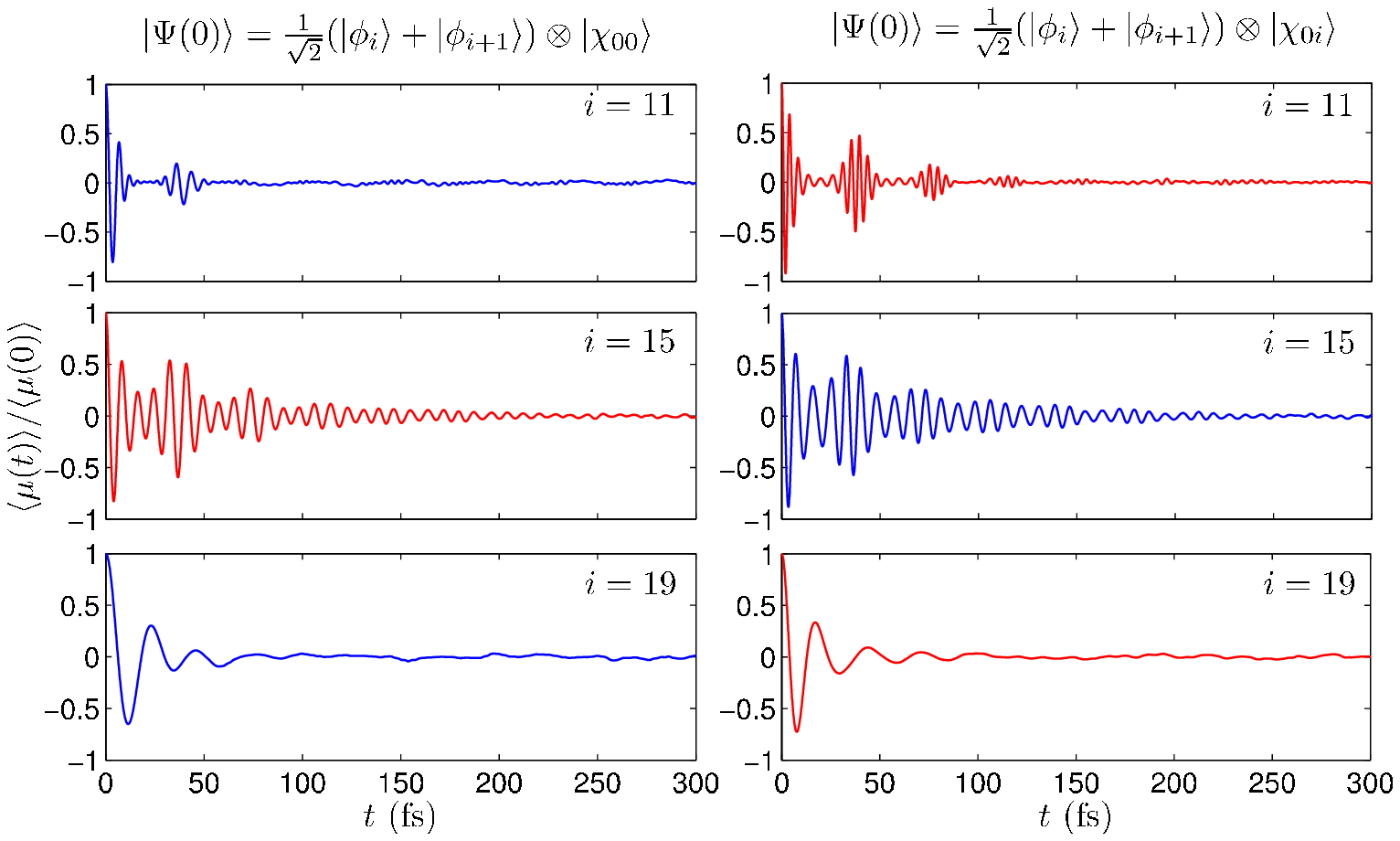}
\caption{Electronic decoherence dynamics in a 20 site chain for different initial superposition between excited states. The panels show the dynamics and decay of the chain polarization starting from a superposition of the form in \eq{eq:super1} (left panels) or  \eq{eq:super2} (right panels).  }
     \label{fig:20decoherence}
\end{figure}

Consider first the case of a chain with 20 CH units. The left panels in \fig{fig:20decoherence} show the time dependence of the polarization when the system is initially prepared in the superposition in \eq{eq:super1} for different $i$'s. The figure shows that the decoherence dynamics can change substantially depending on the  pair of states that are selected to form the superposition. It is even possible to find superpositions for which the coherences are unusually long lived. For example, for  $i=15$ the electronic coherences survive for $\sim 200$ fs,  a timescale that is comparable to the coherence lifetime observed in photosynthetic systems.

\begin{figure}[htbp]
\centering
\includegraphics[width=0.5\textwidth]{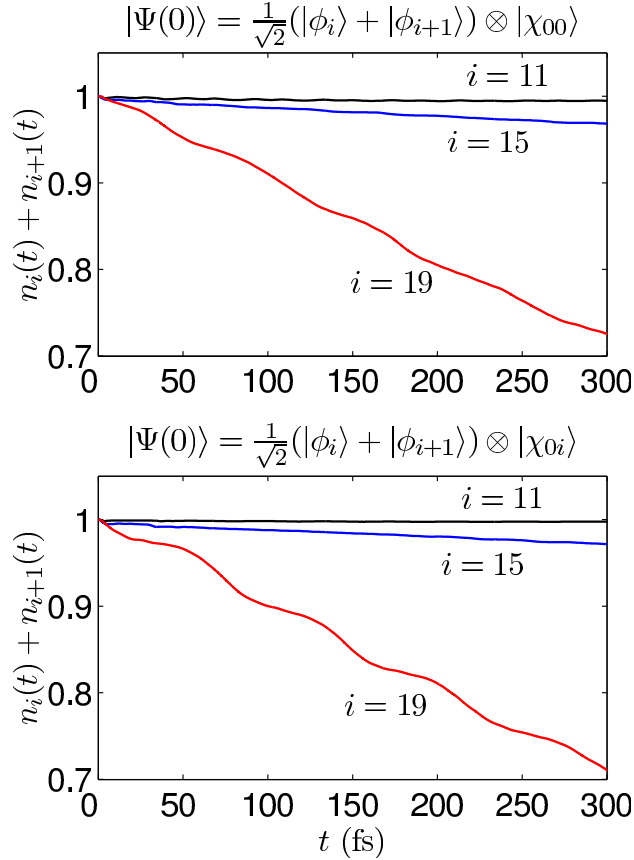}
\caption{Combined population of the $i^{th}$ and $(i+1)^{th}$ orbitals during the decoherence dynamics of a 20-site SSH chain starting from the initial superpositions 
depicted in \fig{fig:decoherence2}. }
     \label{fig:20decoherencepops}
\end{figure}

Additional insights into the decoherence mechanisms are provided by the  shape of $\mean{\hat{\mu}(t)}$ and by the dynamics of population in the excited orbitals  (\fig{fig:20decoherencepops}). The  polarization indicates that for $i=11$ there is vibronic evolution in the excited states that leads to a decay and to the recurrence of the nuclear overlap integrals determining $\mean{\hat{\mu}(t)}$.  By contrast, for $i=19$ this motion is not apparent in $\mean{\hat{\mu}(t)}$ which shows a decay in $\sim 60$ fs with no apparent 
additional structure.  The population dynamics (\fig{fig:20decoherencepops}, upper panel) complements this picture by showing that for $i=11$ a negligible amount of population decays to other levels, while for $i=19$ the transfer of population to other levels is substantial. These observations suggest that for $i=11$ the main mechanism for decoherence is due to anharmonicities in the excited state potential energy surfaces, while for $i=19$ the main decoherence mechanism is due to population transfer to other electronically excited states for which only poor nuclear overlaps are possible.  The case of $i = 15$ is discussed below.

As  an additional test of these observations consider the dynamics of superpositions between the same set of levels but starting from \eq{eq:super2}, that uses a different initial nuclear state. The results are shown in the right panel of \fig{fig:20decoherence}. As can be seen, for $i=11$ changing the initial nuclear state triples the coherence lifetime of the superposition, with three visible recurrences instead of one. Since for this superposition there is negligible amount of population being transferred to other electronic states (see \fig{fig:20decoherencepops}), the data confirms that anharmonicities in the potential energy surfaces of the excited states are the main source of the decoherence in this case.  By contrast, for $i=19$ changing the nature of the initial nuclear state has little effect on the decoherence dynamics, suggesting that the main decoherence mechanism in this case is due to population transfer to other electronic states, as seen in \fig{fig:20decoherencepops}.

The case for $i=15$ where long coherences are observed is different. For this superposition, little population is transferred to other electronic states and a change in the initial nuclear state has little effect on the decoherence dynamics. This suggests that this superposition is protected from decoherence both by the fact that the density of states is such that the two states involved in the superpositions are weakly coupled to other electronic states, and because the sampled potential energy surfaces are less anharmonic than in the other cases considered.

\subsubsection{The case of a 100-site chain}

\begin{figure}[tp]
\centering
\includegraphics[width=1.0\textwidth]{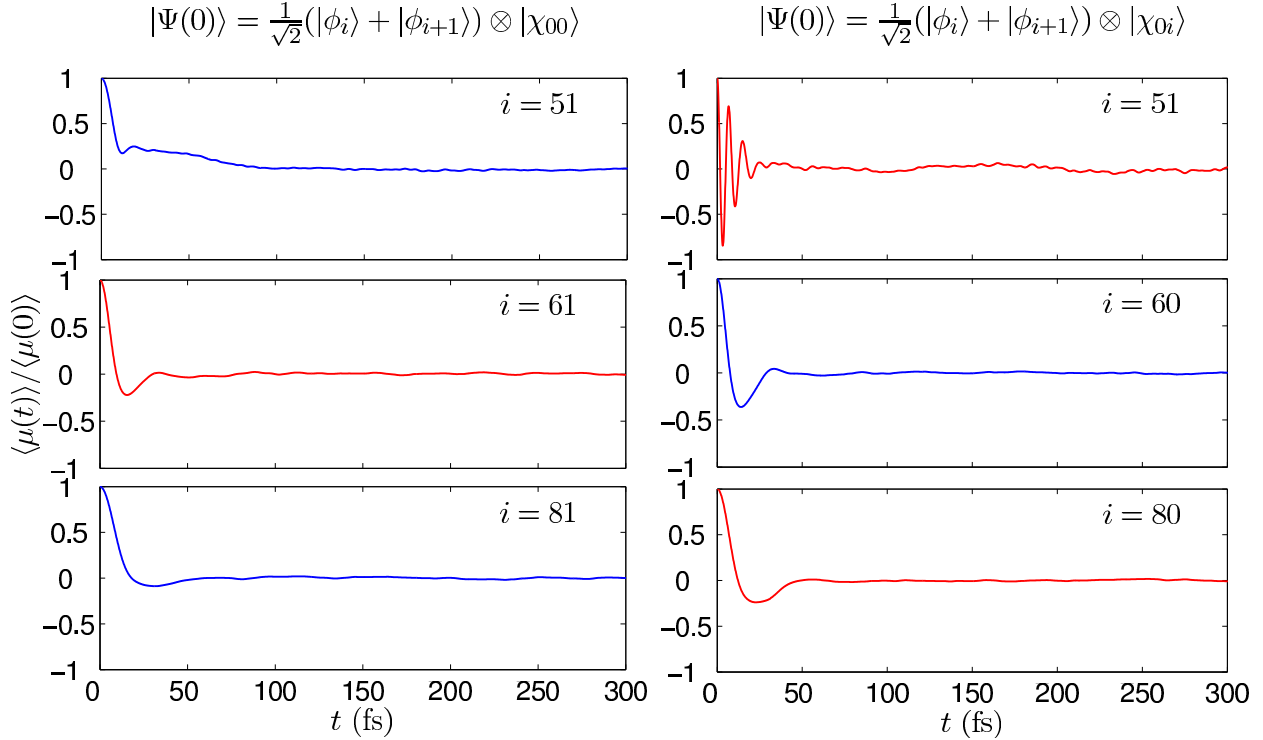}
\caption{Electronic decoherence dynamics in a 100 site chain for different initial superposition states. The panels show the dynamics and decay of the chain polarization starting from a superposition of the form in \eq{eq:super1} (left panels) or  \eq{eq:super2} (right panels).  }
     \label{fig:100decoherence}
\end{figure}

\begin{figure}[htbp]
\centering
\includegraphics[width=0.5\textwidth]{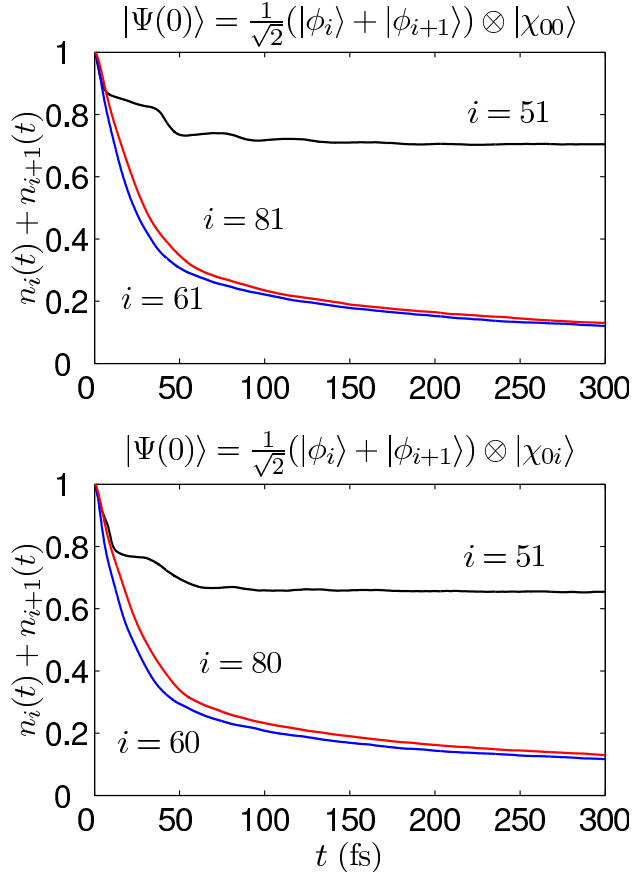}
\caption{Combined population of the $i$th and $(i+1)$-th orbitals during the decoherence dynamics of a 100-site SSH chain starting from the initial superpositions described in \fig{fig:decoherence2}. }
     \label{fig:100decoherencepops}
\end{figure}

For larger systems the situation is qualitatively different. Figure~\ref{fig:100decoherence} shows the dynamics of the polarization for chains initially in a superposition state of the form in \eq{eq:super1} and \eq{eq:super2} for different $i$. Figure~\ref{fig:100decoherencepops} shows the associated change in population of the $i$th and $(i+1)$th orbitals.  The electronic spectrum is so dense  that upon evolution significant population is transferred to other electronic states.  Irrespective of the type of superpositions considered or the initial nuclear state coherence decay in this chain is extremely fast, of the order of 50 fs. The electronic spectrum in this system is simply too dense to maintain electronic coherence. In $\langle\mu (t) \rangle$, however, for $i = 51$ population loss is less than in the other two cases, consistent with the fact that some oscillatory character is visible for $i = 51$ in Fig. \ref{fig:100decoherence}.


\section{Conclusions}

In this paper, we have presented numerical simulations of the electronic coherence  dynamics of PA oligomers of varying length  in which the evolution of both electrons and nuclei are followed explicitly in a mixed quantum-classical approximation. We investigated the 
decoherence of superpositions, as manifest in the dynamics of the 
polarization,  between the ground and excited and between pairs of excited states. 
Decoherence is caused by the decay of the overlap of the nuclear wavepackets associated with all electronic states involved in the superposition. Two basic mechanisms for such decay were identified: population transfer into other electronic states where only poor overlaps are possible, and vibronic evolution in anharmonic potential energy surfaces that lead to wavepacket spread. 

The simulations indicate that for long chains (e.g. $N=100$) the electronic spectrum is so dense that decoherence is dominated by population decay into other states. In this case, no recurrences are observed in the polarization and the decoherence occurs  in tens of  femtoseconds. Further, the decoherence dynamics was found to be largely independent of the type of initial superposition that is subject to the decoherence. By contrast,  for shorter chains (e.g. $N=20$) the simulations indicate that the decoherence dynamics depends strongly on the initial vibronic state. We identified superpositions for which anharmonicities were the main source of decoherence and  superpositions for which population transfer to other electronic states was determinant. Interestingly,  we also observed a  superposition state  between excited states with coherence properties that are long lived, for $\sim$200 fs.  Such a superposition was found to be long lived because it is  spectrally  isolated from other electronic states and because the vibronic dynamics leads to a relatively slow spread of the nuclear wavepackets.

\section{Acknowledgments}  This work was supported by the National Sciences and Engineering Research Council of Canada, and by a grant from the U.S. Air Force Office of Scientific Research under Contract No. FA9550-10-1-0260. I.F. thanks Profs. Mark A. Ratner and George C. Schatz for their support during the preparation of this manuscript.


\begin{thebibliography}{44}
\expandafter\ifx\csname natexlab\endcsname\relax\def\natexlab#1{#1}\fi
\expandafter\ifx\csname bibnamefont\endcsname\relax
  \def\bibnamefont#1{#1}\fi
\expandafter\ifx\csname bibfnamefont\endcsname\relax
  \def\bibfnamefont#1{#1}\fi
\expandafter\ifx\csname citenamefont\endcsname\relax
  \def\citenamefont#1{#1}\fi
\expandafter\ifx\csname url\endcsname\relax
  \def\url#1{\texttt{#1}}\fi
\expandafter\ifx\csname urlprefix\endcsname\relax\def\urlprefix{URL }\fi
\providecommand{\bibinfo}[2]{#2}
\providecommand{\eprint}[2][]{\url{#2}}

\bibitem[{\citenamefont{Franco et~al.}(2008{\natexlab{a}})\citenamefont{Franco,
  Shapiro, and Brumer}}]{franco08}
\bibinfo{author}{\bibfnamefont{I.}~\bibnamefont{Franco}},
  \bibinfo{author}{\bibfnamefont{M.}~\bibnamefont{Shapiro}}, \bibnamefont{and}
  \bibinfo{author}{\bibfnamefont{P.}~\bibnamefont{Brumer}},
  \bibinfo{journal}{J. Chem. Phys.} \textbf{\bibinfo{volume}{128}},
  \bibinfo{pages}{244905} (\bibinfo{year}{2008}{\natexlab{a}}).

\bibitem[{\citenamefont{Levine and Mart{\'\i}nez}(2007)}]{martinez2007}
\bibinfo{author}{\bibfnamefont{B.~G.} \bibnamefont{Levine}} \bibnamefont{and}
  \bibinfo{author}{\bibfnamefont{T.~J.} \bibnamefont{Mart{\'\i}nez}},
  \bibinfo{journal}{Annu. Rev. Phys. Chem.} \textbf{\bibinfo{volume}{58}},
  \bibinfo{pages}{613} (\bibinfo{year}{2007}).

\bibitem[{\citenamefont{Cheng and Fleming}(2009)}]{fleming09}
\bibinfo{author}{\bibfnamefont{Y.-C.} \bibnamefont{Cheng}} \bibnamefont{and}
  \bibinfo{author}{\bibfnamefont{G.~R.} \bibnamefont{Fleming}},
  \bibinfo{journal}{Annu. Rev. Phys. Chem.} \textbf{\bibinfo{volume}{60}},
  \bibinfo{pages}{241} (\bibinfo{year}{2009}).

\bibitem[{\citenamefont{Breuer and Petruccione}(2002)}]{breuer}
\bibinfo{author}{\bibfnamefont{H.~P.} \bibnamefont{Breuer}} \bibnamefont{and}
  \bibinfo{author}{\bibfnamefont{F.}~\bibnamefont{Petruccione}},
  \emph{\bibinfo{title}{The Theory of Open Quantum Systems}}
  (\bibinfo{publisher}{Oxford University Press}, \bibinfo{address}{New York},
  \bibinfo{year}{2002}).

\bibitem[{\citenamefont{Choi et~al.}(2010)\citenamefont{Choi, Risko, Delgado,
  Kim, Br{\'e}das, and Frisbie}}]{choi2010}
\bibinfo{author}{\bibfnamefont{S.~H.} \bibnamefont{Choi}},
  \bibinfo{author}{\bibfnamefont{C.}~\bibnamefont{Risko}},
  \bibinfo{author}{\bibfnamefont{M.~C.~R.} \bibnamefont{Delgado}},
  \bibinfo{author}{\bibfnamefont{B.}~\bibnamefont{Kim}},
  \bibinfo{author}{\bibfnamefont{J.-L.} \bibnamefont{Br{\'e}das}},
  \bibnamefont{and} \bibinfo{author}{\bibfnamefont{C.~D.}
  \bibnamefont{Frisbie}}, \bibinfo{journal}{J. Am. Chem. Soc.}
  \textbf{\bibinfo{volume}{132}}, \bibinfo{pages}{4358} (\bibinfo{year}{2010}).

\bibitem[{\citenamefont{Kapral}(2006)}]{kapral06}
\bibinfo{author}{\bibfnamefont{R.}~\bibnamefont{Kapral}},
  \bibinfo{journal}{Annu. Rev. Phys. Chem.} \textbf{\bibinfo{volume}{57}},
  \bibinfo{pages}{129} (\bibinfo{year}{2006}).

\bibitem[{\citenamefont{Subotnik and Shenvi}(2011)}]{joe2011}
\bibinfo{author}{\bibfnamefont{J.~E.} \bibnamefont{Subotnik}} \bibnamefont{and}
  \bibinfo{author}{\bibfnamefont{N.}~\bibnamefont{Shenvi}},
  \bibinfo{journal}{J. Chem. Phys.} \textbf{\bibinfo{volume}{134}},
  \bibinfo{pages}{244114} (\bibinfo{year}{2011}).

\bibitem[{\citenamefont{Shapiro and Brumer}(2012)}]{pbnewbook}
\bibinfo{author}{\bibfnamefont{M.}~\bibnamefont{Shapiro}} \bibnamefont{and}
  \bibinfo{author}{\bibfnamefont{P.}~\bibnamefont{Brumer}},
  \emph{\bibinfo{title}{Quantum Control of Molecular Processes}}
  (\bibinfo{publisher}{Wiley-VCH}, \bibinfo{address}{Weinheim},
  \bibinfo{year}{2012}).

\bibitem[{\citenamefont{Nielsen and Chuang}(2000)}]{nielsen}
\bibinfo{author}{\bibfnamefont{M.}~\bibnamefont{Nielsen}} \bibnamefont{and}
  \bibinfo{author}{\bibfnamefont{I.}~\bibnamefont{Chuang}},
  \emph{\bibinfo{title}{Quantum Computation and Quantum Information}}
  (\bibinfo{publisher}{Cambridge University Press}, \bibinfo{year}{2000}).

\bibitem[{\citenamefont{Hwang and Rossky}(2004)}]{hwang04}
\bibinfo{author}{\bibfnamefont{H.}~\bibnamefont{Hwang}} \bibnamefont{and}
  \bibinfo{author}{\bibfnamefont{P.~J.} \bibnamefont{Rossky}},
  \bibinfo{journal}{J. Phys. Chem. B} \textbf{\bibinfo{volume}{108}},
  \bibinfo{pages}{6723} (\bibinfo{year}{2004}).

\bibitem[{\citenamefont{Kamisaka et~al.}(2006)\citenamefont{Kamisaka, Kilina,
  Yamashita, and Prezhdo}}]{prezdho06}
\bibinfo{author}{\bibfnamefont{H.}~\bibnamefont{Kamisaka}},
  \bibinfo{author}{\bibfnamefont{S.~V.} \bibnamefont{Kilina}},
  \bibinfo{author}{\bibfnamefont{K.}~\bibnamefont{Yamashita}},
  \bibnamefont{and} \bibinfo{author}{\bibfnamefont{O.~V.}
  \bibnamefont{Prezhdo}}, \bibinfo{journal}{Nano Lett.}
  \textbf{\bibinfo{volume}{6}}, \bibinfo{pages}{2295} (\bibinfo{year}{2006}).

\bibitem[{\citenamefont{Habenicht et~al.}(2007)\citenamefont{Habenicht,
  Kamisaka, Yamashita, and Prezhdo}}]{prezdho07}
\bibinfo{author}{\bibfnamefont{B.~F.} \bibnamefont{Habenicht}},
  \bibinfo{author}{\bibfnamefont{H.}~\bibnamefont{Kamisaka}},
  \bibinfo{author}{\bibfnamefont{K.}~\bibnamefont{Yamashita}},
  \bibnamefont{and} \bibinfo{author}{\bibfnamefont{O.~V.}
  \bibnamefont{Prezhdo}}, \bibinfo{journal}{Nano Lett.}
  \textbf{\bibinfo{volume}{7}}, \bibinfo{pages}{3260} (\bibinfo{year}{2007}).

\bibitem[{\citenamefont{Engel et~al.}(2007)\citenamefont{Engel, Calhoun, Read,
  Ahn, Mancal, Cheng, Blankenship, and Fleming}}]{engel07}
\bibinfo{author}{\bibfnamefont{G.~S.} \bibnamefont{Engel}},
  \bibinfo{author}{\bibfnamefont{T.~R.} \bibnamefont{Calhoun}},
  \bibinfo{author}{\bibfnamefont{E.~L.} \bibnamefont{Read}},
  \bibinfo{author}{\bibfnamefont{T.-K.} \bibnamefont{Ahn}},
  \bibinfo{author}{\bibfnamefont{T.}~\bibnamefont{Mancal}},
  \bibinfo{author}{\bibfnamefont{Y.-C.} \bibnamefont{Cheng}},
  \bibinfo{author}{\bibfnamefont{R.~E.} \bibnamefont{Blankenship}},
  \bibnamefont{and} \bibinfo{author}{\bibfnamefont{G.~R.}
  \bibnamefont{Fleming}}, \bibinfo{journal}{Nature}
  \textbf{\bibinfo{volume}{446}}, \bibinfo{pages}{782} (\bibinfo{year}{2007}).

\bibitem[{\citenamefont{Lee et~al.}(2007)\citenamefont{Lee, Cheng, and
  Fleming}}]{lee07}
\bibinfo{author}{\bibfnamefont{H.}~\bibnamefont{Lee}},
  \bibinfo{author}{\bibfnamefont{Y.-C.} \bibnamefont{Cheng}}, \bibnamefont{and}
  \bibinfo{author}{\bibfnamefont{G.~R.} \bibnamefont{Fleming}},
  \bibinfo{journal}{Science} \textbf{\bibinfo{volume}{316}},
  \bibinfo{pages}{1462} (\bibinfo{year}{2007}).

\bibitem[{\citenamefont{Collini et~al.}(2010)\citenamefont{Collini, Wong, Wilk,
  Curmi, Brumer, and Scholes}}]{collini09}
\bibinfo{author}{\bibfnamefont{E.}~\bibnamefont{Collini}},
  \bibinfo{author}{\bibfnamefont{C.~Y.} \bibnamefont{Wong}},
  \bibinfo{author}{\bibfnamefont{K.~E.} \bibnamefont{Wilk}},
  \bibinfo{author}{\bibfnamefont{P.~M.~G.} \bibnamefont{Curmi}},
  \bibinfo{author}{\bibfnamefont{P.}~\bibnamefont{Brumer}}, \bibnamefont{and}
  \bibinfo{author}{\bibfnamefont{G.~D.} \bibnamefont{Scholes}},
  \bibinfo{journal}{Nature} \textbf{\bibinfo{volume}{463}},
  \bibinfo{pages}{644} (\bibinfo{year}{2010}).

\bibitem[{\citenamefont{Mohseni et~al.}(2008)\citenamefont{Mohseni, Rebentrost,
  Lloyd, and Aspuru-Guzik}}]{Mohseni2008}
\bibinfo{author}{\bibfnamefont{M.}~\bibnamefont{Mohseni}},
  \bibinfo{author}{\bibfnamefont{P.}~\bibnamefont{Rebentrost}},
  \bibinfo{author}{\bibfnamefont{S.}~\bibnamefont{Lloyd}}, \bibnamefont{and}
  \bibinfo{author}{\bibfnamefont{A.}~\bibnamefont{Aspuru-Guzik}},
  \bibinfo{journal}{J. Chem. Phys.} \textbf{\bibinfo{volume}{129}},
  \bibinfo{pages}{174106} (\bibinfo{year}{2008}).

\bibitem[{\citenamefont{Lloyd}(2009)}]{Lloyd2009}
\bibinfo{author}{\bibfnamefont{S.}~\bibnamefont{Lloyd}},
  \bibinfo{journal}{Nature Phys} \textbf{\bibinfo{volume}{5}},
  \bibinfo{pages}{164} (\bibinfo{year}{2009}).

\bibitem[{\citenamefont{Ishizaki and Fleming}(2009{\natexlab{a}})}]{akihito09}
\bibinfo{author}{\bibfnamefont{A.}~\bibnamefont{Ishizaki}} \bibnamefont{and}
  \bibinfo{author}{\bibfnamefont{G.~R.} \bibnamefont{Fleming}},
  \bibinfo{journal}{Proc. Natl. Acad. Sci. USA} \textbf{\bibinfo{volume}{106}},
  \bibinfo{pages}{17255} (\bibinfo{year}{2009}{\natexlab{a}}).

\bibitem[{\citenamefont{Briggs and Eisfeld}(2011)}]{briggs2011}
\bibinfo{author}{\bibfnamefont{J.~S.} \bibnamefont{Briggs}} \bibnamefont{and}
  \bibinfo{author}{\bibfnamefont{A.}~\bibnamefont{Eisfeld}},
  \bibinfo{journal}{Phys. Rev. E} \textbf{\bibinfo{volume}{83}},
  \bibinfo{pages}{051911} (\bibinfo{year}{2011}).

\bibitem[{\citenamefont{Katz et~al.}(2008)\citenamefont{Katz, Gelman, Ratner,
  and Kosloff}}]{Katz2008}
\bibinfo{author}{\bibfnamefont{G.}~\bibnamefont{Katz}},
  \bibinfo{author}{\bibfnamefont{D.}~\bibnamefont{Gelman}},
  \bibinfo{author}{\bibfnamefont{M.~A.} \bibnamefont{Ratner}},
  \bibnamefont{and} \bibinfo{author}{\bibfnamefont{R.}~\bibnamefont{Kosloff}},
  \bibinfo{journal}{J. Chem. Phys.} \textbf{\bibinfo{volume}{129}},
  \bibinfo{pages}{034108} (\bibinfo{year}{2008}).

\bibitem[{\citenamefont{Renaud et~al.}(2011)\citenamefont{Renaud, Ratner, and
  Mujica}}]{nico2011}
\bibinfo{author}{\bibfnamefont{N.}~\bibnamefont{Renaud}},
  \bibinfo{author}{\bibfnamefont{M.~A.} \bibnamefont{Ratner}},
  \bibnamefont{and} \bibinfo{author}{\bibfnamefont{V.}~\bibnamefont{Mujica}},
  \bibinfo{journal}{J. Chem. Phys.} \textbf{\bibinfo{volume}{135}},
  \bibinfo{pages}{075102} (\bibinfo{year}{2011}).

\bibitem[{\citenamefont{Ishizaki and
  Fleming}(2009{\natexlab{b}})}]{Ishizaki2009}
\bibinfo{author}{\bibfnamefont{A.}~\bibnamefont{Ishizaki}} \bibnamefont{and}
  \bibinfo{author}{\bibfnamefont{G.~R.} \bibnamefont{Fleming}},
  \bibinfo{journal}{J. Chem. Phys.} \textbf{\bibinfo{volume}{130}},
  \bibinfo{pages}{234111} (\bibinfo{year}{2009}{\natexlab{b}}).

\bibitem[{\citenamefont{Kelly and Rhee}(2011)}]{kelly2011}
\bibinfo{author}{\bibfnamefont{A.}~\bibnamefont{Kelly}} \bibnamefont{and}
  \bibinfo{author}{\bibfnamefont{Y.~M.} \bibnamefont{Rhee}},
  \bibinfo{journal}{J. Phys. Chem. Lett.} \textbf{\bibinfo{volume}{2}},
  \bibinfo{pages}{808} (\bibinfo{year}{2011}).

\bibitem[{\citenamefont{Pachon and Brumer}(2011)}]{leonardo}
\bibinfo{author}{\bibfnamefont{L.}~\bibnamefont{Pachon}} \bibnamefont{and}
  \bibinfo{author}{\bibfnamefont{P.}~\bibnamefont{Brumer}},
  \bibinfo{journal}{J. Phys. Chem. Lett.} \textbf{\bibinfo{volume}{2}},
  \bibinfo{pages}{2728} (\bibinfo{year}{2011}).

\bibitem[{\citenamefont{Pachon and Brumer}(submitted)}]{pachonpccp}
\bibinfo{author}{\bibfnamefont{L.}~\bibnamefont{Pachon}} \bibnamefont{and}
  \bibinfo{author}{\bibfnamefont{P.}~\bibnamefont{Brumer}},
  \bibinfo{journal}{Phys. Chem. Chem. Phys.}  (\bibinfo{year}{submitted}).

\bibitem[{\citenamefont{Huo and Coker}(2011)}]{coker}
\bibinfo{author}{\bibfnamefont{P.}~\bibnamefont{Huo}} \bibnamefont{and}
  \bibinfo{author}{\bibfnamefont{D.}~\bibnamefont{Coker}}, \bibinfo{journal}{J.
  Phys. Chem. Lett.} \textbf{\bibinfo{volume}{2}}, \bibinfo{pages}{825}
  (\bibinfo{year}{2011}).

\bibitem[{\citenamefont{Sterpone et~al.}(2011)\citenamefont{Sterpone,
  Martinazzo, Panda, and Burghardt}}]{burghardt2011}
\bibinfo{author}{\bibfnamefont{F.}~\bibnamefont{Sterpone}},
  \bibinfo{author}{\bibfnamefont{R.}~\bibnamefont{Martinazzo}},
  \bibinfo{author}{\bibfnamefont{A.~N.} \bibnamefont{Panda}}, \bibnamefont{and}
  \bibinfo{author}{\bibfnamefont{I.}~\bibnamefont{Burghardt}},
  \bibinfo{journal}{Z. Phys. Chem.} \textbf{\bibinfo{volume}{225}},
  \bibinfo{pages}{541} (\bibinfo{year}{2011}).

\bibitem[{\citenamefont{Heeger et~al.}(1988)\citenamefont{Heeger, Kivelson,
  Schrieffer, and Su}}]{SSH}
\bibinfo{author}{\bibfnamefont{A.~J.} \bibnamefont{Heeger}},
  \bibinfo{author}{\bibfnamefont{S.}~\bibnamefont{Kivelson}},
  \bibinfo{author}{\bibfnamefont{J.~R.} \bibnamefont{Schrieffer}},
  \bibnamefont{and} \bibinfo{author}{\bibfnamefont{W.~P.} \bibnamefont{Su}},
  \bibinfo{journal}{Rev. Mod. Phys.} \textbf{\bibinfo{volume}{60}},
  \bibinfo{pages}{781} (\bibinfo{year}{1988}).

\bibitem[{\citenamefont{Teramoto et~al.}(2009)\citenamefont{Teramoto, Wang,
  Kobryanskii, Taneichi, and Kobayashi}}]{kobayashi2009}
\bibinfo{author}{\bibfnamefont{T.}~\bibnamefont{Teramoto}},
  \bibinfo{author}{\bibfnamefont{Z.}~\bibnamefont{Wang}},
  \bibinfo{author}{\bibfnamefont{V.~M.} \bibnamefont{Kobryanskii}},
  \bibinfo{author}{\bibfnamefont{T.}~\bibnamefont{Taneichi}}, \bibnamefont{and}
  \bibinfo{author}{\bibfnamefont{T.}~\bibnamefont{Kobayashi}},
  \bibinfo{journal}{Phys. Rev. B} \textbf{\bibinfo{volume}{79}},
  \bibinfo{pages}{033202} (\bibinfo{year}{2009}).

\bibitem[{\citenamefont{Adachi et~al.}(2002)\citenamefont{Adachi, Kobryanskii,
  and Kobayashi}}]{kobayashi2002}
\bibinfo{author}{\bibfnamefont{S.}~\bibnamefont{Adachi}},
  \bibinfo{author}{\bibfnamefont{V.~M.} \bibnamefont{Kobryanskii}},
  \bibnamefont{and}
  \bibinfo{author}{\bibfnamefont{T.}~\bibnamefont{Kobayashi}},
  \bibinfo{journal}{Phys. Rev. Lett.} \textbf{\bibinfo{volume}{89}},
  \bibinfo{pages}{027401} (\bibinfo{year}{2002}).

\bibitem[{\citenamefont{Tretiak et~al.}(2003)\citenamefont{Tretiak, Saxena,
  Martin, and Bishop}}]{sergei2003}
\bibinfo{author}{\bibfnamefont{S.}~\bibnamefont{Tretiak}},
  \bibinfo{author}{\bibfnamefont{A.}~\bibnamefont{Saxena}},
  \bibinfo{author}{\bibfnamefont{R.~L.} \bibnamefont{Martin}},
  \bibnamefont{and} \bibinfo{author}{\bibfnamefont{A.~R.}
  \bibnamefont{Bishop}}, \bibinfo{journal}{Proc. Natl. Acad. Sci. USA}
  \textbf{\bibinfo{volume}{100}}, \bibinfo{pages}{2185} (\bibinfo{year}{2003}).

\bibitem[{\citenamefont{Stella et~al.}(2011)\citenamefont{Stella, Miranda,
  Horsfield, and Fisher}}]{stella2011}
\bibinfo{author}{\bibfnamefont{L.}~\bibnamefont{Stella}},
  \bibinfo{author}{\bibfnamefont{R.~P.} \bibnamefont{Miranda}},
  \bibinfo{author}{\bibfnamefont{A.~P.} \bibnamefont{Horsfield}},
  \bibnamefont{and} \bibinfo{author}{\bibfnamefont{A.~J.}
  \bibnamefont{Fisher}}, \bibinfo{journal}{J. Chem. Phys.}
  \textbf{\bibinfo{volume}{134}}, \bibinfo{pages}{194105}
  (\bibinfo{year}{2011}).

\bibitem[{\citenamefont{Ness and Fisher}(1999)}]{ness1999}
\bibinfo{author}{\bibfnamefont{H.}~\bibnamefont{Ness}} \bibnamefont{and}
  \bibinfo{author}{\bibfnamefont{A.~J.} \bibnamefont{Fisher}},
  \bibinfo{journal}{Phys. Rev. Lett.} \textbf{\bibinfo{volume}{83}},
  \bibinfo{pages}{452} (\bibinfo{year}{1999}).

\bibitem[{\citenamefont{Tully}(1998)}]{tully}
\bibinfo{author}{\bibfnamefont{J.~C.} \bibnamefont{Tully}}, in
  \emph{\bibinfo{booktitle}{Classical and Quantum Dynamics in Condensed Phase
  Simulations}}, edited by
  \bibinfo{editor}{\bibfnamefont{B.}~\bibnamefont{Berne}},
  \bibinfo{editor}{\bibfnamefont{G.}~\bibnamefont{Ciccotti}}, \bibnamefont{and}
  \bibinfo{editor}{\bibfnamefont{D.~F.} \bibnamefont{Coker}}
  (\bibinfo{publisher}{World Scientific}, \bibinfo{address}{Singapore},
  \bibinfo{year}{1998}), pp. \bibinfo{pages}{700--720}.

\bibitem[{\citenamefont{Streitwolf}(1998)}]{streitwolf1}
\bibinfo{author}{\bibfnamefont{H.~W.} \bibnamefont{Streitwolf}},
  \bibinfo{journal}{Phys. Rev. B} \textbf{\bibinfo{volume}{58}},
  \bibinfo{pages}{14356} (\bibinfo{year}{1998}).

\bibitem[{\citenamefont{Johansson and Stafstr\"om}(2002)}]{johansson1}
\bibinfo{author}{\bibfnamefont{A.}~\bibnamefont{Johansson}} \bibnamefont{and}
  \bibinfo{author}{\bibfnamefont{S.}~\bibnamefont{Stafstr\"om}},
  \bibinfo{journal}{Phys. Rev. B} \textbf{\bibinfo{volume}{65}},
  \bibinfo{pages}{045207} (\bibinfo{year}{2002}).

\bibitem[{\citenamefont{Franco et~al.}(2008{\natexlab{b}})\citenamefont{Franco,
  Shapiro, and Brumer}}]{francowire}
\bibinfo{author}{\bibfnamefont{I.}~\bibnamefont{Franco}},
  \bibinfo{author}{\bibfnamefont{M.}~\bibnamefont{Shapiro}}, \bibnamefont{and}
  \bibinfo{author}{\bibfnamefont{P.}~\bibnamefont{Brumer}},
  \bibinfo{journal}{J. Chem. Phys.} \textbf{\bibinfo{volume}{128}},
  \bibinfo{pages}{244906} (\bibinfo{year}{2008}{\natexlab{b}}).

\bibitem[{\citenamefont{Franco et~al.}(2007)\citenamefont{Franco, Shapiro, and
  Brumer}}]{francowireprl}
\bibinfo{author}{\bibfnamefont{I.}~\bibnamefont{Franco}},
  \bibinfo{author}{\bibfnamefont{M.}~\bibnamefont{Shapiro}}, \bibnamefont{and}
  \bibinfo{author}{\bibfnamefont{P.}~\bibnamefont{Brumer}},
  \bibinfo{journal}{Phys. Rev. Lett.} \textbf{\bibinfo{volume}{99}},
  \bibinfo{pages}{126802} (\bibinfo{year}{2007}).

\bibitem[{\citenamefont{Hillery et~al.}(1984)\citenamefont{Hillery, O'Connell,
  Scully, and Wigner}}]{wigner}
\bibinfo{author}{\bibfnamefont{M.}~\bibnamefont{Hillery}},
  \bibinfo{author}{\bibfnamefont{R.~F.} \bibnamefont{O'Connell}},
  \bibinfo{author}{\bibfnamefont{M.~O.} \bibnamefont{Scully}},
  \bibnamefont{and} \bibinfo{author}{\bibfnamefont{E.~P.}
  \bibnamefont{Wigner}}, \bibinfo{journal}{Phys. Rep.}
  \textbf{\bibinfo{volume}{106}}, \bibinfo{pages}{121} (\bibinfo{year}{1984}).

\bibitem[{\citenamefont{Chao and Wang}(1985)}]{phonon1}
\bibinfo{author}{\bibfnamefont{K.~A.} \bibnamefont{Chao}} \bibnamefont{and}
  \bibinfo{author}{\bibfnamefont{Y.}~\bibnamefont{Wang}}, \bibinfo{journal}{J.
  Phys. C:Solid State Phys.} \textbf{\bibinfo{volume}{18}},
  \bibinfo{pages}{L1127} (\bibinfo{year}{1985}).

\bibitem[{\citenamefont{Mele and Rice}(1980)}]{phonon2}
\bibinfo{author}{\bibfnamefont{E.~J.} \bibnamefont{Mele}} \bibnamefont{and}
  \bibinfo{author}{\bibfnamefont{M.~J.} \bibnamefont{Rice}},
  \bibinfo{journal}{Phys. Rev. Lett.} \textbf{\bibinfo{volume}{45}},
  \bibinfo{pages}{926} (\bibinfo{year}{1980}).

\bibitem[{\citenamefont{Wilson et~al.}(1980)\citenamefont{Wilson, Decius, and
  Cross}}]{wilson}
\bibinfo{author}{\bibfnamefont{E.~B.} \bibnamefont{Wilson}},
  \bibinfo{author}{\bibfnamefont{J.~C.} \bibnamefont{Decius}},
  \bibnamefont{and} \bibinfo{author}{\bibfnamefont{P.~C.} \bibnamefont{Cross}},
  \emph{\bibinfo{title}{Molecular Vibrations}} (\bibinfo{publisher}{Dover},
  \bibinfo{address}{New York}, \bibinfo{year}{1980}).

\bibitem[{\citenamefont{Schlosshauer}(2008)}]{schlosshauer}
\bibinfo{author}{\bibfnamefont{M.}~\bibnamefont{Schlosshauer}},
  \emph{\bibinfo{title}{Decoherence and the Quantum-to-Classical Transition}}
  (\bibinfo{publisher}{Springer}, \bibinfo{address}{New York},
  \bibinfo{year}{2008}).

\bibitem[{\citenamefont{Franco and Brumer}(2008)}]{francosym}
\bibinfo{author}{\bibfnamefont{I.}~\bibnamefont{Franco}} \bibnamefont{and}
  \bibinfo{author}{\bibfnamefont{P.}~\bibnamefont{Brumer}},
  \bibinfo{journal}{J. Phys. B} \textbf{\bibinfo{volume}{41}},
  \bibinfo{pages}{074003} (\bibinfo{year}{2008}).

\end{thebibliography}

\end{document}